\documentclass{elsart}

\usepackage{graphicx}
\usepackage{tabularx}
\usepackage{array}
\usepackage{amssymb,amsmath}
\usepackage{lineno}

\begin{document}


\begin{frontmatter}

\title{A new approach to study energy-dependent arrival delays on
photons from astrophysical sources}

\author{Manel Mart\'{i}nez\corauthref{cor1}} and
\ead{martinez@ifae.es}
\author{Manel Errando}

\address{Institut de F\'{i}sica d'Altes Energies,
Edifici Cn Universitat Auton\`{o}ma de Barcelona,  08193
Bellaterra, Spain}

\corauth[cor1]{Corresponding author. Tel. +34 935811309, Fax +34 935811938.}

\begin{abstract}
Correlations between the arrival time and the energy of photons
emitted in outbursts of astrophysical objects are predicted in
quantum and classical gravity scenarios and can appear as well as
a result of complex acceleration mechanisms responsible for the
photon emission at the source. This paper presents a robust method
to study such correlations that overcomes some limitations
encountered in previous analysis, and is based on a Likelihood
function built from the physical picture assumed for the emission,
propagation and detection of the photons. The results of the
application of this method to a flare of Markarian 501 observed by
the MAGIC telescope are presented. The method is also applied to a
simulated dataset based on the flare of \mbox{PKS 2155-304}
recorded by the H.E.S.S. observatory to proof its applicability to
complex photon arrival time distributions.
\end{abstract}

\end{frontmatter}

\section{Introduction}
\label{intro}

Observations of the electromagnetic radiation from astrophysical
sources exhibiting fast flux variations --mainly active galactic nuclei (AGNs) and gamma-ray bursts (GRBs) and pulsars-- make possible the study of
correlations between the energy and the arrival time of individual
photons. Two different effects can explain the appearance of such
correlations: either they are source-intrinsic or have originated
during the propagation of the radiation from the source to the
Earth. On the one hand, the emission mechanisms taking place at
the source can cause delays between photons of different energies
to appear. In the case of pulsars, a difference in the emission location for different energies can cause the shape and peak time of the pulses to change with energy \cite{cheng,harding}. In AGNs the delays can be caused, for instance, by the acceleration of the electrons responsible for the emission \cite{ghisellini,bednarek07}. On the other hand, if the
emission is assumed to be simultaneous at the source for all the
photons in a certain energy range, an energy-dependent propagation
effect could also explain the appearance of these correlations.
Quantum Gravity (QG) \cite{rovelli,ellis92,kostelecky,amelino97,gambini} and classical gravitation --through the existence of wormholes
\cite{klinkhamer07}-- are two frameworks
where this kind of propagation effects may show up.

It is widely speculated that space-time is a dynamical medium,
subject to quantum-gravitational effects that cause space-time to
fluctuate on the Planck distance scales. Model realizations of QG
predict that quantum fluctuations in the space-time metric make it
appear `foamy' on very short distance scales (see
\cite{rovelli98,Ellis:1999yd,sarkar02,piran05} for reviews). The propagation of photons
through this fluctuating space-time might induce an energy
dependence of the speed of light, resulting in an observable
difference on the propagation time for photons of different
energies. The anomaly induced by QG will always be a small
perturbation of the assumed light speed $c$, and can therefore be
treated as a correction of the form \mbox{$\Delta c / c = - E /
M_{QG1}$} or \mbox{$-E^2 / M_{QG2}^2 $} where $E$ is the photon
energy, $M_{QG1}$ and $M_{QG2}$ are respectively the effective QG
scales in the linear and the quadratic term, and the minus sign
indicates that most of the models expect this effect to be
subluminal.

The possibility to observe QG efects in the propagation of photons from astrophysical sources was proposed by Amelino-Camelia et al. \cite{amelino98}. If two photons of different energy are emitted simultaneously from
an astrophysical object, the expected delay between their arrival
times when they are detected increases with the distance to the
source and with the energy of the photons. Therefore, the maximum
sensitivity to energy dependent propagation effects is expected
from observations of very fast flux variations coming from sources
at large distances that emit photons up to very high energies.

The first candidates considered for such observations were
gamma-ray bursts (GRB) \cite{amelino98}, but the first experimental result on possible
energy-dependent speed of light came from the measurement of a flare of
the active galactic nucleus (AGN) Mkn 421 at TeV energies by the Whipple
gamma-ray telescope \cite{biller99},
 claiming
a lower limit to the energy scale of quantum gravity of
\mbox{$M_{QG1}
> 6 \times 10^{16}$} GeV
. Other bounds have been
obtained by studying the emission of pulsars \cite{kaaret99}, and
a combined analysis of many GRBs yielded to a robust lower limit
of $M_{QG1}
> 0.9 \times 10^{16}$\:GeV \cite{ellis06}.

The highest sensitivity of the new generation of ground-based
gamma-ray telescopes was expected to have a clear impact in these
studies \cite{blanch03} and, indeed, has provided new observations
of fast flares of AGNs at TeV energies with richer photon
statistics than the previous generation of instruments could
\cite{magic-flare,aharonian07}, enhancing the sensitivity of these
measurements to any energy dependent propagation effect.

This paper presents a new method of analysis that can be applied
to any observed set of photons to search for any kind of
correlations between their arrival time and energy. This analysis
technique is described in section \:\ref{method} and applied to
the Mkn 501 flare observed by the MAGIC telescope in July 2005
\cite{magic-flare} in section \:\ref{application-MAGIC}. An
application to a simulated set of photons based on the more
complex flare structure of the PKS 2155-304 outburst observed by
the H.E.S.S. collaboration in July 2006 \cite{aharonian07} is
discussed in section \:\ref{application-HESS}. Finally, the
conclusions of this work are summarized in section
\:\ref{conclusions}.

\section{Description of the method}
\label{method}

The pioneering search for QG efects using TeV gamma-ray data by
the Whipple Collaboration \cite{biller99} was based on few tens of
gamma-ray events that where both binned in time and energy.

Given the fact that the statistics for sources at the largest
distances from us and photons of the highest energies (conditions
that maximize the expected photon delay) is usually scarce, in
order to make an optimal use of the information of the arrival
time and estimated energy of each recorded photon the analysis
method used should be unbinned.

Some analysis methods, such as the Modified Cross Correlation
Function (MCCF) \cite{li,edelson} or the Continuous Wavelet
Transform (CWT) \cite{mallat} have been widely used for similar
analysis \cite{ellis06,ellis03,lamon,bolmont,HESS-QG}. These
methods have been applied using the information of the arrival
time for each individual photon without binning, but binning the
energy in bands in order to determine the time lag as a function
of the energy and therefore did not make optimal use of the photon
information.

In addition, as mentioned before, measurements at the highest
photon energies maximize the expected delays between photons, but
all the sources detected at the GeV-TeV energy range show steeply
falling energy spectra, reducing the photon statistics as the
energy increases. Some of the above analysis methods, as for
instance the wavelet approach, \cite{ellis03} simply cannot deal
properly with the low photon statistics obtained in most of the
very high energy gamma-ray measurements.

Completely unbinned approaches to search for correlations in
similar data sets are nevertheless possible. For instance the
Energy Cost Function (ECF) method described in \cite{magic-QG} has
been successfully used to measure a tiny time lag in the same data
set which will be used here as an example. Nevertheless, that
method requires the identification of well-isolated flares and
cannot be applied directly to complex lightcurves where several
overlapping flares, or simply other kind of features such as edges
of non-contained flares, appear in the lightcurves.

This work describes a completely unbinned method which overcomes
the above limitations and can be applied in a general manner to
any data set, regardless on the photon statistics and the
lightcurve shape and structure. Moreover, it makes use of an
optimal estimator which, if properly applied, would use all the
information contained in the data and therefore provide with the
most accurate estimate of the correlation between energy and
arrival time. This estimator is the Likelihood Function built from
the combined probability of having observed a set of photons with
individual energy $E_i$ arriving at time $t_i$, and has to be
constructed from a physical description of the assumed processes
involved in the emission, propagation and detection of the
photons.

Any hypotheses about the physics mechanisms that finally produce
the measured set of photons can be accommodated in this analysis
provided a reasonably simple mathematical description is at hand.
A possible general physics picture of the process is the
following:

\begin{itemize}
\item{} Photons were produced at the source following an intrinsic
lightcurve\footnote{The term ``lightcurve'' is widely used in
astrophysics to denote the photon emission time distribution
function.} $F_s(t_s)$ where $t_s$ is the time at the source and an
intrinsic energy spectrum $\Gamma(E_s)$ where $E_s$ is the energy
of the photons at the source. It might be that already at the
source the lighcurve depends on energy or the energy spectrum
depends on time. If models describing these dependencies are at
hand for the observed astrophysical source, they can be readily
incorporated in the approach described here.

\item{} These photons propagate in space and there, if there is an
energy-dependent refraction index related to some effective energy
scale parameter $M_{QGn}$, a delay $D(E_s,M_{QGn},z)$ should
affect their propagation time.

\item{} These photons reach the Earth and are recorded by satellite
or ground-based
instruments as photons with measured energy $E$ and arrival time
$t$. These detectors have typically an extremely good time
resolution compared with the flare time scales but a limited
energy resolution $G(E-E_s,\sigma_E(E_s))$ which may be any
complex function of the observed energy.
\end{itemize}

\subsection{Probability Density Function}

A mathematical expression casting the probability density function
(p.d.f.) describing the physics picture given above is the
following:

\begin{equation}
\frac{dP}{dE \, dt} = N \, \int_0^{\infty} \, {\Gamma(E_s) \, C(E_s,t) \,
G(E-E_s,\sigma_E(E_s)) \, F_s(t-D(E_s,M_{QGn},z)) \, d E_s }
\label{likelihood-function}
\end{equation}

where

\begin{itemize}

\item $P$ is the probability density function (p.d.f.) for a
photon to have observed energy $E$ and arrival time $t$.

\item $\Gamma(E_s)$ is the photon energy distribution at the
source. For instance, for Very High Energy gamma rays, in most
cases the energy spectrum is well described in first approximation
by a pure power law \mbox{$\Gamma(E_s) = \Gamma_0 \cdot
(E_s/E_0)^{-\alpha}$} where $\alpha$ is the differential spectral
index. Nevertheless, more complex functions such as broken or
curved power laws, spectral index for the photons coming from the
flare activity different from the spectral index for the photons
coming from the steady activity, or synchrotron or inverse Compton
spectra can be readily used. The spectrum can be obtained from a
fit to the whole photon dataset assuming that it does not change
during the flare. Otherwise, if the photon distribution is
observed to change with time it can be fit to the data assuming a
specific mechanism for correlation at the source.

\item $C(E_s,t)$ accounts for changes in the effective area of
the detector during the observation of the flare. In most cases
this factor will be constant and can be neglected. It has to be
included, for example, if the data comes from ground-based
telescopes with changing observation conditions (significant
changes in the zenith angle of the pointing, atmospheric conditions,
etc.) or from satellites observing in survey mode.

\item $G(E-E_s,\sigma_E(E_s))$ is the gamma energy smearing
produced by the instrument. In first approximation it could be a
gaussian distribution with width $\sigma_E(E_s)$ although more
complex functions can also be used. This function has to be
obtained from the study of the energy response of the detector.

\item $D(E_s,M_{QGn},z)$ is the eventual propagation delay as a
function of the photon energy $E_s$, an effective energy scale
$M_{QGn}$ and the source redshift $z$. This function may
accommodate linear, quadratic or any arbitrary realization of any
possible propagation delay mechanisms. A general implementation of
this function in terms of an effective expansion in the case of
Quantum Gravity scenarios is discussed below.

\item $t_s =t-D(E_s,M_{QGn},z)$ is the actual photon production
time at the source obtained from the measured arrival time
corrected by the delay generated by non-trivial propagation
effects.

\item $F_s(t_s)$ is the emission time distribution of the photons
at the source. For most astrophysical sources there are no compelling
models which can be used to predict and/or fit
the flaring lightcurves. Moreover, since the probability $P$
depends only on the relative delay of the photons for different
energies, the best estimator for the time
structure at the source can be taken to be the one of the observed
gamma emission time distribution for the lowest energy gamma rays.
In the specific case of Very High Energy gamma rays, this
estimator is a very good choice because, since the energy spectra
are typically steep power laws, most of the recorded gammas are
collected at the lowest energies, and therefore the gamma
statistics is then sufficient to infer the lightcurve as explained
above.

\item $N$ is the p.d.f. normalization. This normalization is
extremely important for the proper use of the probability in a
likelihood function. To have an unbiased estimation of the
fitting parameters it has to be computed in such a way that the
p.d.f. integral is equal to 1 no matter the values of the free
parameters in the p.d.f.

\end{itemize}

If the propagation delays are inspired in Quantum Gravity models,
in a power $n$ realization of Quantum gravity effects ($n=1$
linear and $n=2$ quadratic) for $z<<1$ we have that the
propagation delay of a photon of energy $E_s$ with respect to the
arrival time of a photon of energy $E_{s,0}$ is:

\begin{equation}
 D(E_s,M_{QGn},z) = \frac{E_s^n -E_{s,0}^n}{M_{QGn}^n} \frac{z}{H_0}
\end{equation}

where $z$ is the source redshift and $H_0$ is the local Hubble
constant (in inverse seconds). For sources at larger distances the
redshift in the photon energies and the expansion effect in the
photon path have to be taken into account, and then the correct
expressions are:

\begin{equation}
 D(E_s,E_{QG},z) = \frac{1}{H_0} \frac{E_s -E_{s,0}}{M_{QG1}}
 \int_0^z \frac{(1+z) dz}{h(z)}
\end{equation}

for the linear case and

\begin{equation}
 D(E_s,E_{QG},z) = \frac{1}{H_0}
 \frac{E_s^2 -E_{s,0}^2} {M_{QG2}^2}
 \int_0^z \frac{(1+z)^{2} dz}{h(z)}
\end{equation}

for the quadratic case where

\begin{equation}
h(z) = \sqrt{ \Omega_{\Lambda} + \Omega_M (1+z)^3 }
\end{equation}

being $\Omega_{\Lambda}$ and $\Omega_M$ the standard cosmological
parameters.

\subsection{Likelihood Function and fitting procedure}

\par With the above described p.d.f. the likelihood function for
the observation of all the $N_{\gamma}$ photons recorded during
the flare can be build as

\begin{equation}
L = \prod_1^{N_{\gamma}} \frac{dP}{dE_i \, dt_i}
\end{equation}

\par  In the above expression, several free parameters may appear.
One of them is obviously the effective scale $M_{QGn}$ at which
propagation delays appear. Other parameters may try to fit the
intrinsic lightcurve shape to some analytic or semi-analytic
parametrization or the intrinsic energy spectrum. The likelihood
function can be fitted to the data using a numerical procedure to
find the set of free parameters that minimize the function $-2 \,
log(L)$.

From the conceptual point of view, when the $M_{QGn}$ parameter is
changed, the effect is basically sliding in time the lightcurve
profile as a function of energy to fit the possible correlation
between energy and arrival time in the data. If there is no
correlation between energy and time, or the correlation is not
significative that would correspond to $M_{QGn} \rightarrow \pm
\infty$ and therefore $M_{QGn}$ is not a good fitting parameter
for the expected small correlations, because the likelihood
function would be highly non-parabolic. Instead, a well-behaved
fitting parameter is $\hat{M}_P/M_{QGn}$, where $\hat{M}_P = 2.4
\times 10^{18}$ GeV is the reduced Planck mass, which can have
positive or negative or zero values and will likely show a nice
parabolic behavior. Moreover, in most QG scenarios its order of
magnitude is expected to be around 1 which makes numerical
calculations easier.

In the physics picture assumed, when sliding in time the
lightcurve all gammas are expected to lead to the same ``intrinsic''
lightcurve, therefore all gammas, regardless on their energy,
contribute in the fitting procedure to either determine or test
the assumed lightcurve shape even if just the lower energy ones
were initially used to estimate the shape of the lightcurve.

If the fit is well behaved, the fit results for all free
parameters shall show a parabolic behavior of the log-likelihood
function around the ``best fit minimum''. Around that minimum, an
increase of  \mbox{$\Delta (-2 \, log(L))= 1$} will then give the
single-parameter one-$\sigma$ uncertainties.

Since the Likelihood p.d.f. by construction has dimensions
$[E]^{-1}[t]^{-1}$, its value at the minimum does not have a
direct interpretation and does not provide a simple means to
estimate the goodness-of-fit for the assumed function. Therefore
additional tests must be performed to quantify the adequacy of the
physics picture casted in the Likelihood function to describe the
data.

One simple possibility that we have used is the following: once
the Likelihood fit is done, the data and the Likelihood function
computed at the best fit parameter values are binned in energy and
time. Then the $\chi^2$ value of the comparison of the resulting
bi-dimensional histograms (in energy and time) is computed and
compared with the number of degrees of freedom, providing a
goodness-of-fit probability. This approach allows also an easy
visualization on the adequacy of the fitting function to describe
the data. Other possibilities such as integrating the likelihood
function would require the use of complex and lengthy
multi-dimensional integrations.

\section{Example of application to a Markarian 501 flare}
\label{application-MAGIC}

The blazar Markarian 501, located at $z=0.034$, showed a very high
activity in the gamma-ray band in June and July 2006. On July 9 a
very strong flare in the TeV regime was observed by the MAGIC
collaboration \cite{magic-flare}. The analysis procedure explained
in the previous section was applied to this set of photons in
reference \cite{magic-QG} together with a method based on
maximizing the total energy in the most active part of the flare.

\begin{figure}
\centering
\includegraphics[width=0.9\textwidth]{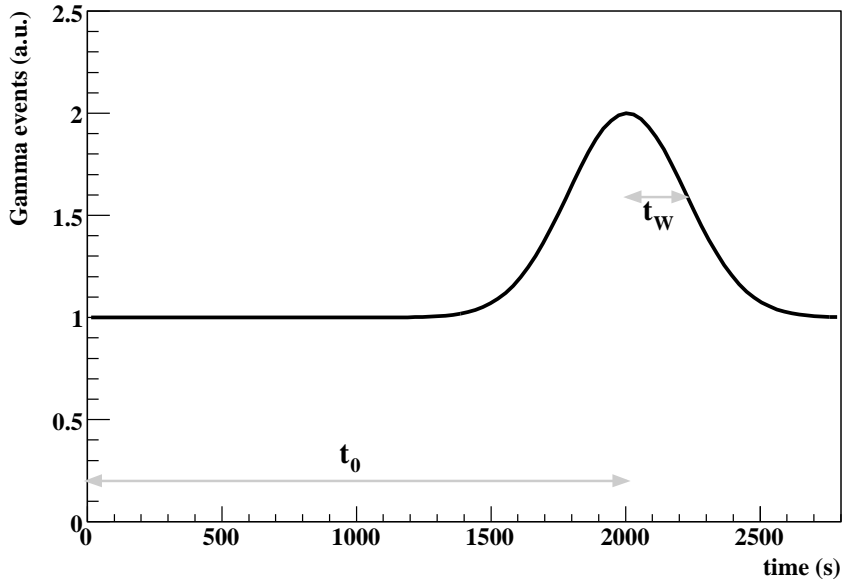}
\caption{Parameterized flare shape for the Markarian 501 MAGIC
flare data: the data are well described by a Gaussian-shaped flare
activity on top of a constant baseline of steady gamma-ray
emission (called "baseline" elsewere in this work).}
\label{fig:MAGIC-parameterization}
\end{figure}

In that data sample, the lightcurve was tested to be well
described by a simple gaussian flare on top of a baseline constant
in time within the observation window
(see Fig.~\ref{fig:MAGIC-parameterization}). It was tested that other
mathematical functions fitting also well the data flare shape,
such as the halving-doubling time formula used in the original
publication \cite{magic-flare} or even a simple triangle function,
were producing similar final results. Therefore, the time
distribution was parameterized as a gaussian flare of width $t_W$
and position $t_0$ relative to the first gamma arrival time, on
top of a flat background time distribution (see figure
\ref{fig:MAGIC-parameterization}).

The intrinsic spectrum was obtained from a global fit to the flare
data and consisted of two power-laws $\Gamma(E_s) = (E_s/1 \,
TeV)^{-\alpha}$ with different spectral indices: $\alpha = 2.4$
for the gaussian excess component and $\alpha = 2.7$ for the
continuous baseline \cite{magic-flare}. These values were changed
within few standard deviations without a significant impact on the
final results.

The energy resolution was assumed to be linear with energy
\mbox{$\sigma_E = 0.22 \times E$} although several other more complex
energy dependencies consistent with the MAGIC telescope energy
resolution measurements were tested with no significant impact in
the results.

To infer from the data the best estimators for the fitting
parameters, the likelihood function was fitted to the data using a
standard numerical minimization package to minimize the function
$-2 \, log(L)$ as a function of four free parameters:

\begin{itemize}
\item $\hat{M}_P/M_{QGn}$ the effective energy scale for a
linear or a quadratic Quantum Gravity realization.

\item $t_W$ the gaussian intrinsic flare width.

\item $t_0$ the maximum intrinsic flare position relative to the
first gamma arrival time.

\item $x_B$ the normalization in area between the flat baseline
component and the superimposed gaussian flare.
\end{itemize}

\begin{figure}
\centering
\includegraphics[width=0.9\textwidth]{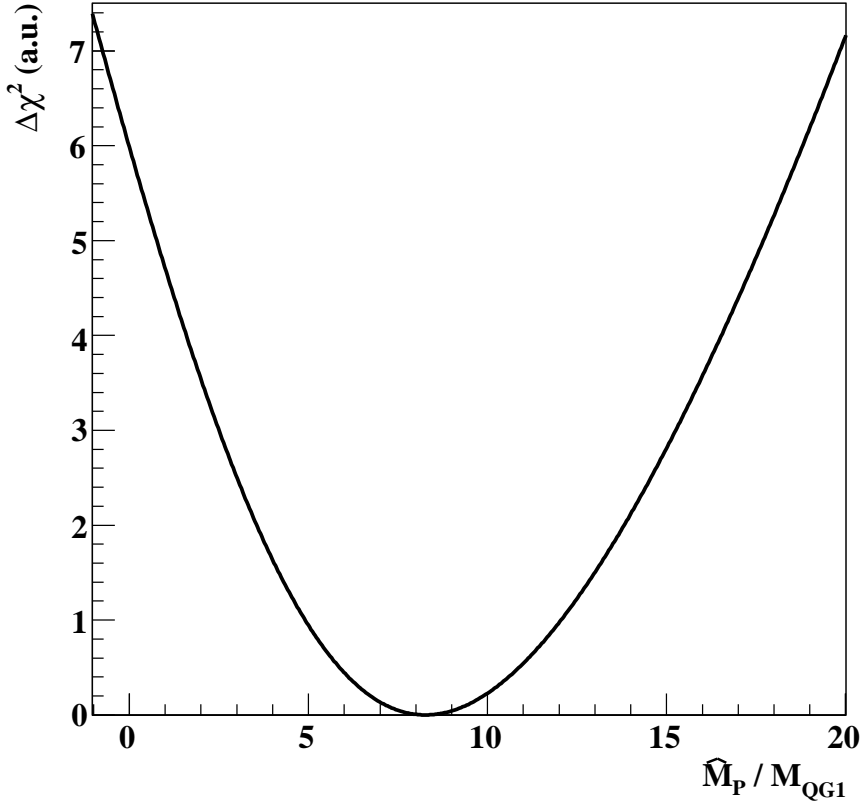}
\caption{Behavior of the function $-2 \, log(L)$ with respect to
the parameter $\hat{M}_P/M_{QG1}$ around the minimum. The shape is
rather parabolic and allows excluding the value $\hat{M}_P/M_{QG1}
= 0$ at more than 2 sigma significance ( $\Delta (-2 \, log(L))=
4$ ).} \label{fig:MAGIC-likelihood-parabolic}
\end{figure}

The fit converged smoothly towards a minimum at \mbox{$M_{QG1} =
0.30^{+0.24}_{- 0.10} \times 10^{18}$} GeV for the linear case and
\mbox{$M_{QG2} = 0.57^{+ 0.75}_{- 0.19} \times 10^{11}$} GeV for
the quadratic one. The uncertainties given above are asymmetric
because they are given on $M_{QGn}$ and not on
$\hat{M}_P/M_{QGn}$, the actual fitting parameter, and
correspond to the change in the fitting parameter leading to an
increase of $\Delta (- 2 \, log(L))= 1$ around the minimum. The
shape of the Likelihood function as a function of
$\hat{M}_P/M_{QG1}$ around that minimum is almost parabolic as can
be gleaned from Fig.~\ref{fig:MAGIC-likelihood-parabolic} and
therefore, the above uncertainty can be considered as enclosing
one-sigma single parameter statistical uncertainty. Therefore, the
fitted value of $\hat{M}_P/M_{QG1}$ differs from zero by over
two-sigma.

The values of the fitting parameters in the linear and quadratic
scan are shown in Tab.~\ref{tab:fit}. As discussed in reference
\cite{magic-QG} this result is perfectly consistent with the one
obtained by using the Energy Cost Function method described there.
The correlation between the different parameters is always below
50\% except between $M_{QGn}$ and $t_{0}$, which are correlated at
a level of 60\% as the first parameter carries the information
about the delay introduced by a QG effect during the propagation
of photons and the second indicates where the maximum of the flare
is located.

\begin{table}
        \centering
\begin{tabular}{ccc}
\hline\hline
    & $n=1$& $n=2$\\
   \hline
    $\hat{M}_P/M_{QGn}$ & $8.9^{+5.8}_{-4.5}$ &
      $4.3^{+2.7}_{-1.7} \times 10^{7}$ \\
    $t_{W}$ & $219^{+28}_{-29}$ s & $\left(228 \pm 27\right)$ s \\
    $t_{0}$ & $\left(2005 \pm 42 \right)$ s & $\left(2041 \pm 36\right)$ s \\
    $x_{B}$ & $0.38 \pm 0.04$ & $0.38 \pm 0.04$\\
    \hline
\end{tabular}
\caption{Best fit values of the free parameters in the Likelihood
function for the linear and the quadratic scan. These values are
obtained when all the parameters of the fit are left free, while
the $M_{QGn}$ values quoted in the text are obtained fixing all
the parameters except for $\hat{M}_P/M_{QGn}$ itself.}
\label{tab:fit}

\end{table}

The goal of this work is just the discussion of the new method
described here and therefore we shall not discuss here the meaning
or the interpretation of the above results. That is properly
discussed in reference \cite{magic-QG}.

Moreover, a series of tests were performed to test the robustness,
the stability and the interpretation of the result.

First, the dataset was fitted to the same Likelihood function
after scrambling randomly the arrival time of the photons. The fit
then produced a best fit value compatible with zero energy-time
correlation at one sigma.

Second, mock simulated Monte Carlo datasets of the size of the
actual data but with input parameters unknown by us were
produced\footnote{These datasets were independently produced by
Adrian Biland with a toy Monte Carlo
simulation implementing his own simple description of the physical
process.} and then fitted blindly by our Likelihood function,
recovering the input parameters within the expected uncertainties.
The same happened with the Energy Cost Function method as
discussed in reference \cite{magic-QG}.

\begin{figure}
\centering
\includegraphics[width=0.9\textwidth]{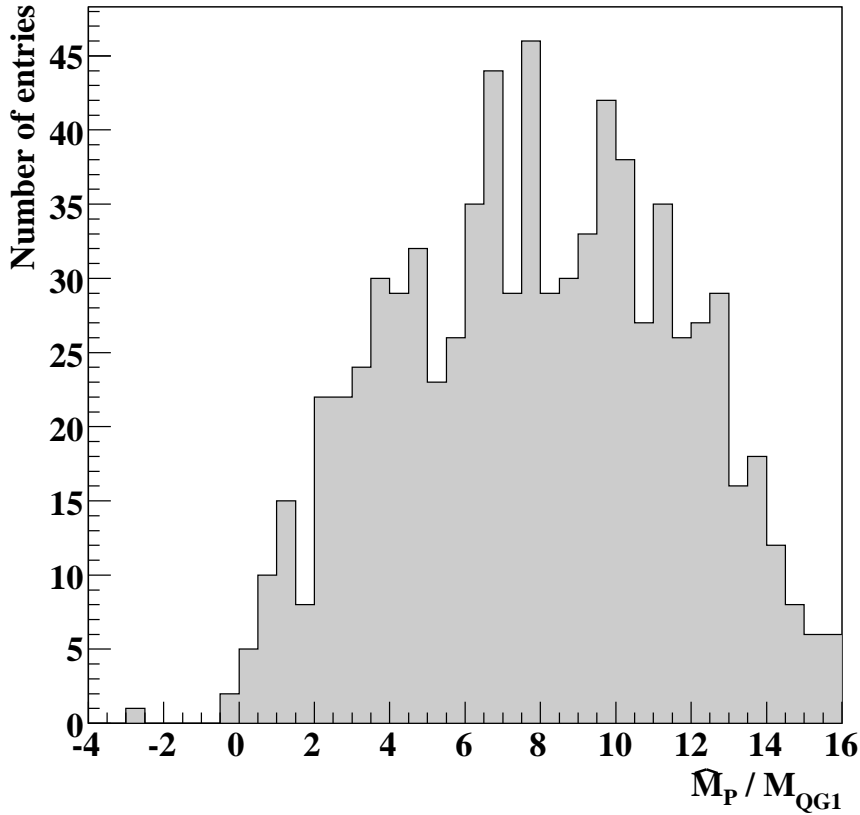}
\caption{Distribution for the best fit parameter
$\hat{M}_P/M_{QG1}$ in bootstrap replica of the MAGIC
flare data. As can be seen, just few fits produce
$\hat{M}_P/M_{QG1} \leq 0$. } \label{fig:MAGIC-bootstrap-plot}
\end{figure}

Third, thousand datasets of the same characteristics of the actual
data were produced using a bootstrap method on the actual data.
Each of these datasets was fitted to our Likelihood function and
the distribution of the best-fit values was studied (see figure
\ref{fig:MAGIC-bootstrap-plot}). As expected from a probabilistic
interpretation of the statistical uncertainties obtained in the
Likelihood fit to the MAGIC data, the
\mbox{$M_{QG1} = 0.30^{+0.24}_{- 0.10} \times 10^{18}$} GeV
was enclosing about 68\% of the best-fit
values and less than 2\% of the best-fit values were below zero.

Therefore, it can be concluded that the Likelihood method
explained here was producing in this case a reliable result.
Moreover, reasonable modifications on the actual details of the
physical process description casted in our Likelihood formula, as
already discussed before, change the results in just a fraction of
the statistical uncertainty.

\begin{figure}
\centering
\includegraphics[width=0.8\textwidth]{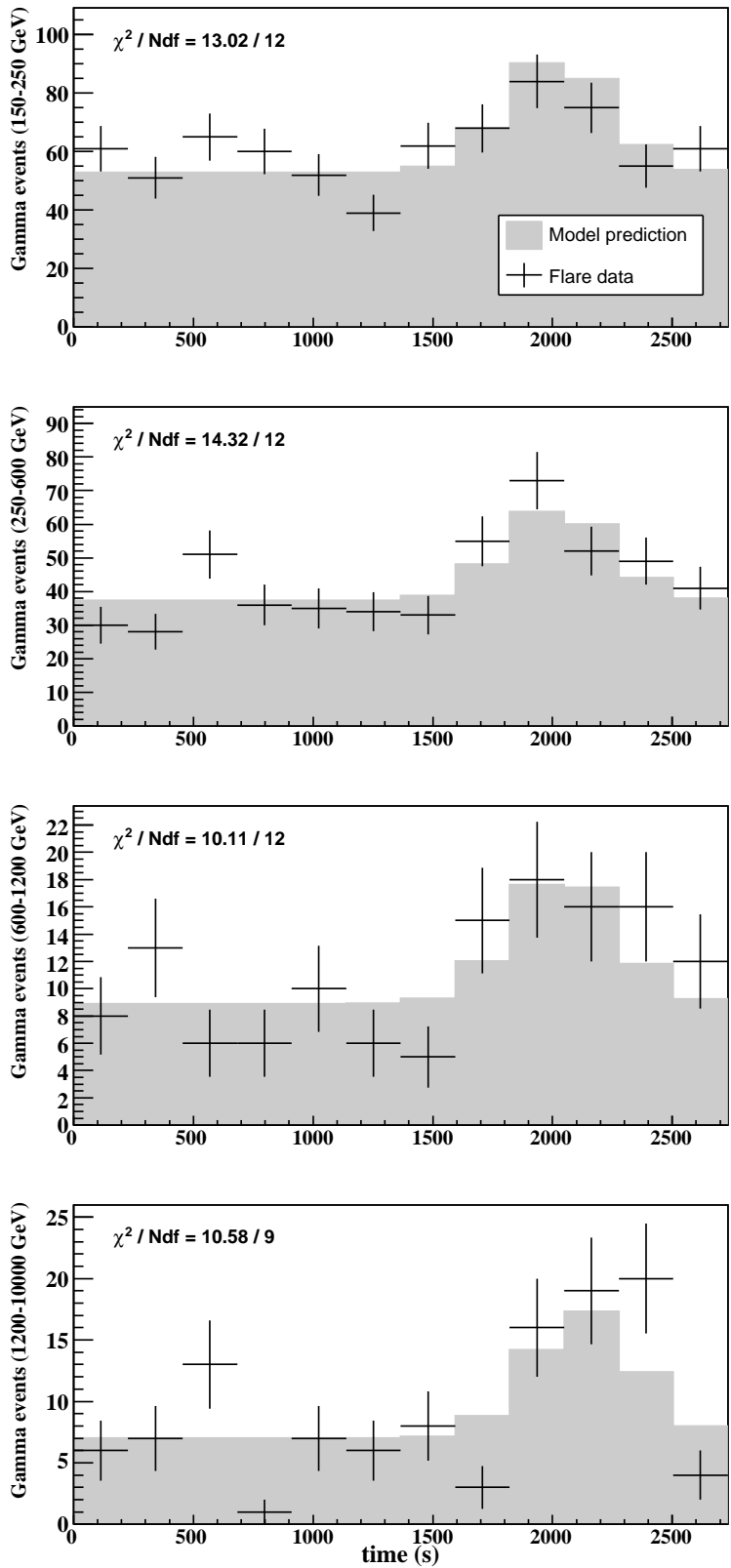}
\caption{Comparison of the binned MAGIC data and Likelihood
function distributions to check the goodness-of-fit.
The overall $\chi^2$/Ndf is $48.0/41$.} \label{fig:MAGIC-binned}
\end{figure}

Finally, in order to check the goodness-of-fit of our Likelihood
formula to the data, both the Likelihood function and the data
have been binned following the binning originally used in
reference \cite{magic-flare}, namely: four bins in energy and
twelve bins in time (see Fig.~\ref{fig:MAGIC-binned}). The
$\chi^2$ obtained by comparing the data and fitting function
histograms turns out to be in reasonable agreement with the number
of degrees of freedom (equal to the number of used bins minus the
number of fitting parameters). Therefore, as can be gleaned from
Fig.~\ref{fig:MAGIC-binned}, our Likelihood function is indeed a
good description of the MAGIC flare data.

\section{Example of application to simulated photons following
a more complex flare structure}
\label{application-HESS}
The H.E.S.S. collaboration published in
year 2007 the observation of a huge flare of the
blazar PKS 2155-304 \cite{aharonian07}, located at  $z=0.116$.
The flare happened during the night of July 26th 2006,
lasted about 90 minutes and was so intense (over 10 thousand gamma
rays collected corresponding to a significance of almost 160
sigma) that a lightcurve with 1-minute time binning was produced.

\begin{figure}
\centering
\includegraphics[width=0.9\textwidth]{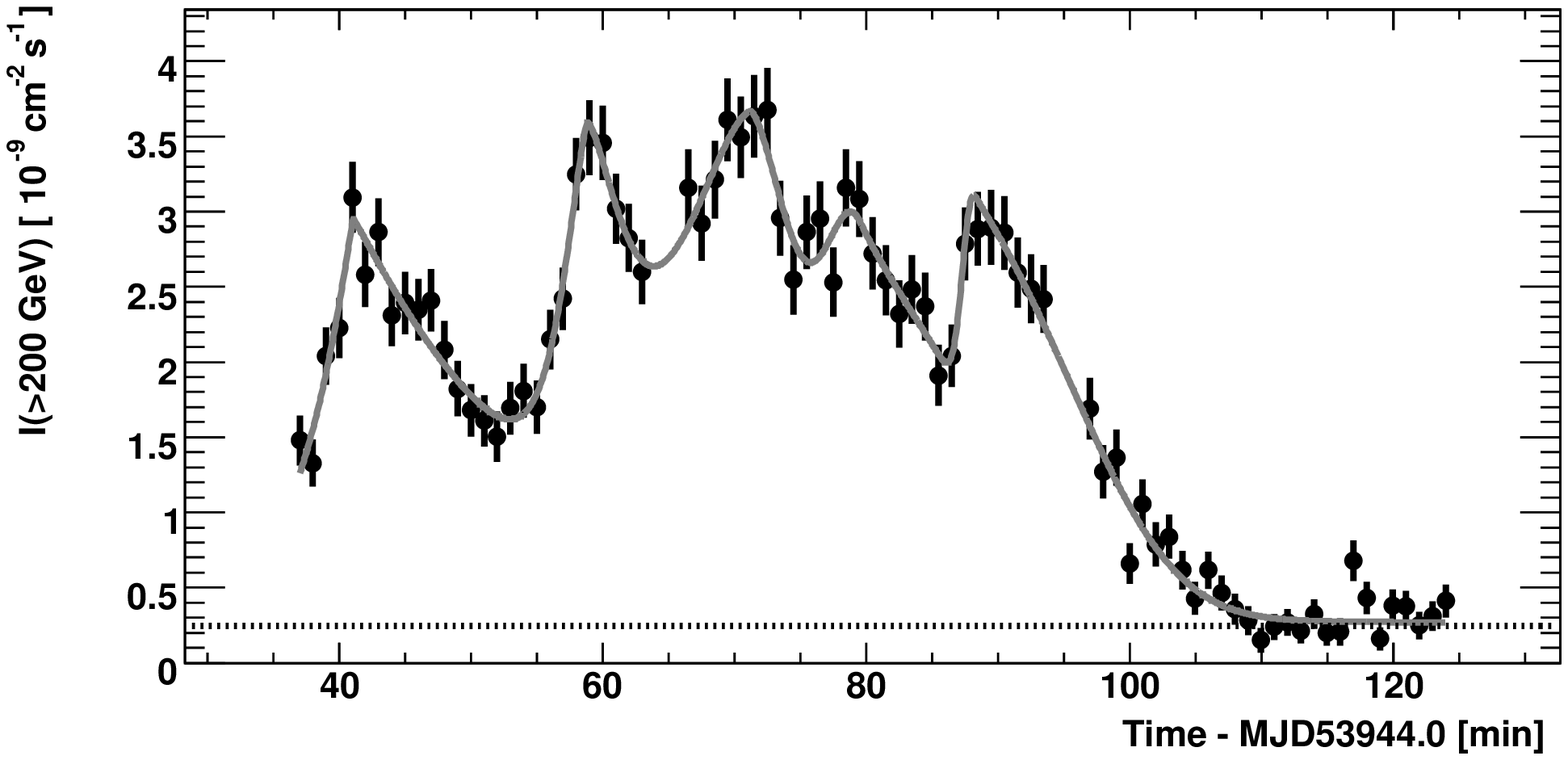}
\caption{The lightcurve observed from the July 26th 2006 flare of
PKS2155-304 by the H.E.S.S. collaboration (fig 1 in reference
\cite{aharonian07}). The data are binned in 1-minute intervals.
The line is the fit to these data of the superposition of five
bursts described in the text.} \label{fig:HESS-lightcurve}
\end{figure}

\begin{table}
        \centering
\begin{tabular}{ccccc}
\hline\hline
$t_{\rm max}$  & $A$ &$\tau_{\rm r}$ & $\tau_{\rm d}$  & $\kappa$ \\
$[$min$]$ & $[10^{-9}\,{\rm cm}^{-2}\,{\rm
    s}^{-1}]$ & [s] & [s] & \\
\hline\\
41.0 & 2.7$\pm$0.2 & 173$\pm$28 & 610$\pm$129 & 1.07$\pm$0.20\\
58.8 & 2.1$\pm$0.9  & 116$\pm$53 & 178$\pm$146 & 1.43$\pm$0.83\\
71.3 & 3.1$\pm$0.3 & 404$\pm$219 & 269$\pm$158 & 1.59$\pm$0.42\\
79.5 & 2.0$\pm$0.8 & 178$\pm$55  & 657$\pm$268 & 2.01$\pm$0.87\\
88.3 & 1.5$\pm$0.5 & 67$\pm$44   & 620$\pm$75  & 2.44$\pm$0.41\\
\hline\\
\end{tabular}
\caption{The results of the best $\chi^2$ fit of the superposition
of five bursts and a constant to the
data shown in Figure~\ref{fig:HESS-lightcurve}. The constant term is
$0.27\pm0.03 \times 10^{-9}\,{\rm cm}^{-2}\,{\rm s}^{-1}$ (1.1
${\rm I}_{\rm Crab}$). Taken from \cite{aharonian07}.\label{burst_info}}
\end{table}

That lightcurve, shown in Fig.~\ref{fig:HESS-lightcurve},
exhibits a rather complex structure in time which, as discussed in
reference \cite{aharonian07}, is well fitted by a set of five
overlapping flares described by using ``generalized Gaussian''
shapes following the formulation suggested by Norris et al.
\cite{norris96}, that is:

$I(t)= A \exp[-(|t-t_{max}|/\sigma_{r,d})^{\kappa}]$

where $\sigma_{r}$ (rise) is to be used for $t < t_{max}$ and
$\sigma_{d}$ (decay) is to be used for $t\geq t_{max}$. The actual
best fit values for $A,\sigma_{r,d}$ and $\kappa$ for each one of
the five overlapping flares are collected in Tab.~\ref{burst_info}.

\begin{figure}
\centering
\includegraphics[width=0.9\textwidth]{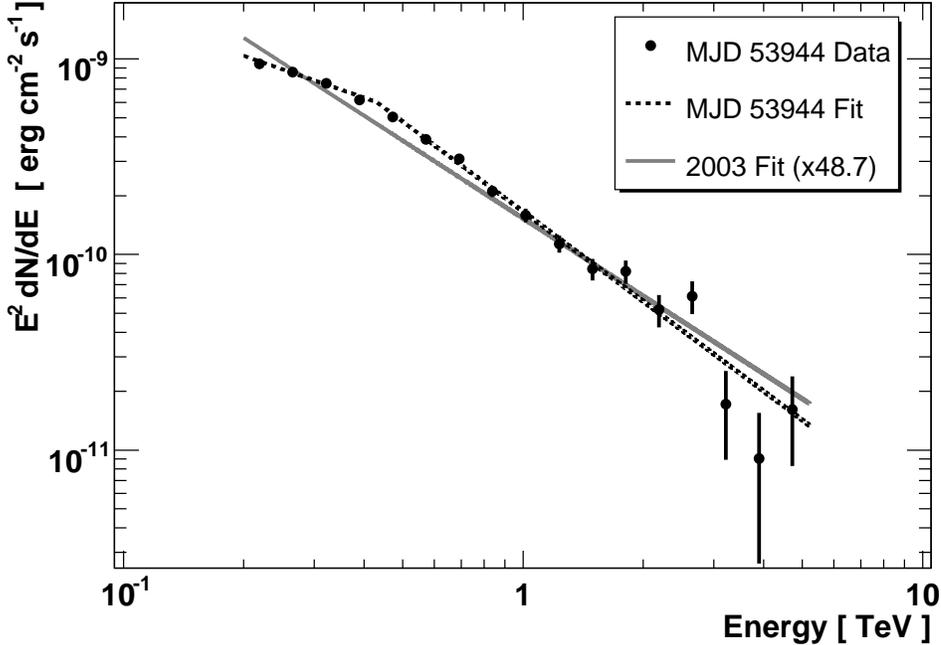}
\caption{The time-averaged differential energy spectrum observed
from the July 26th 2006 flare of PKS 2155-304 by the H.E.S.S.
collaboration (fig 3 in reference \cite{aharonian07}). The dashed
line is the best fit of a broken power-law to the data as
described in the text.} \label{fig:HESS-spectrum}
\end{figure}

In addition, such a huge statistics allowed a precise analysis of
the observed differential energy spectrum which, as can be seen in
Fig.~\ref{fig:HESS-spectrum} turns out to be well described
between about 200 GeV and 5 TeV by a broken power law such as:

\begin{eqnarray}
E < E_B & : & \frac{dN}{dE}= I_0 \large( \frac{E}{1 \,
TeV}\large)^{-{\Gamma}_1} \nonumber \\
E > E_B & : & \frac{dN}{dE}= I_0 \large( \frac{E}{1 \,
TeV}\large)^{{\Gamma}_2-{\Gamma}_1} \, \large( \frac{E}{1 \,
TeV}\large)^{-{\Gamma}_2} \nonumber
\end{eqnarray}

where $I_0 = (2.06 \pm 0.16 \pm 0.41) \times 10^{-10} \, cm^{-2}
s^{-1} TeV^{-1}$, $E_B=(430 \pm 22 \pm 80)$ GeV, ${\Gamma}_1=2.71
\pm 0.06 \pm 0.10$ and ${\Gamma}_2=3.53 \pm 0.05 \pm 0.10$ where
for each parameter the two uncertainties are the statistical and
the systematic values respectively. This energy spectrum has not
been corrected for the absorption of VHE gammas on the
Extragalactic Background Light but corresponds to the observed
one, which is precisely the input needed for the correlation
study.

We wanted to test whether our method could be applied to such a
kind of complex flare to eventually obtain results on any possible
energy-dependent propagation delay effect. Nevertheless, since the
method presented here relies on the use of the individual photon
information and we have no access the actual H.E.S.S. data, Monte
Carlo generated samples of simulated gamma-ray data following the
H.E.S.S. published lightcurve and differential energy spectrum
have been produced to try to understand the actual sensitivity our
method would provide in spotting any non-trivial energy-time
correlation, would it be present in the real data.

For that, Monte Carlo simulated gamma-ray datasets with the same
amount of gammas observed by H.E.S.S., their integral lightcurve,
their differential energy spectrum and their claimed energy
resolution were produced. In each dataset a different value for
$M_{QGn}$ was assumed: from no energy-time correlation
($\hat{M}_P/M_{QGn} = 0$) to the same value found in the fits to
the MAGIC data described before, and in the linear and in the
quadratic scenarios. Those datasets were fitted to a Likelihood
function built following equation \ref{likelihood-function} in
which the functions $\Gamma(E_s)$ , $G(E-E_s,\sigma_E(E_s))$  and
$F_s(t-D(E_s,M_{QGn},z))$ are the same ones used for generating
the simulated datasets, namely, the ones describing the H.E.S.S.
data as discussed above.

\begin{figure}
\centering
\includegraphics[width=0.8\textwidth]{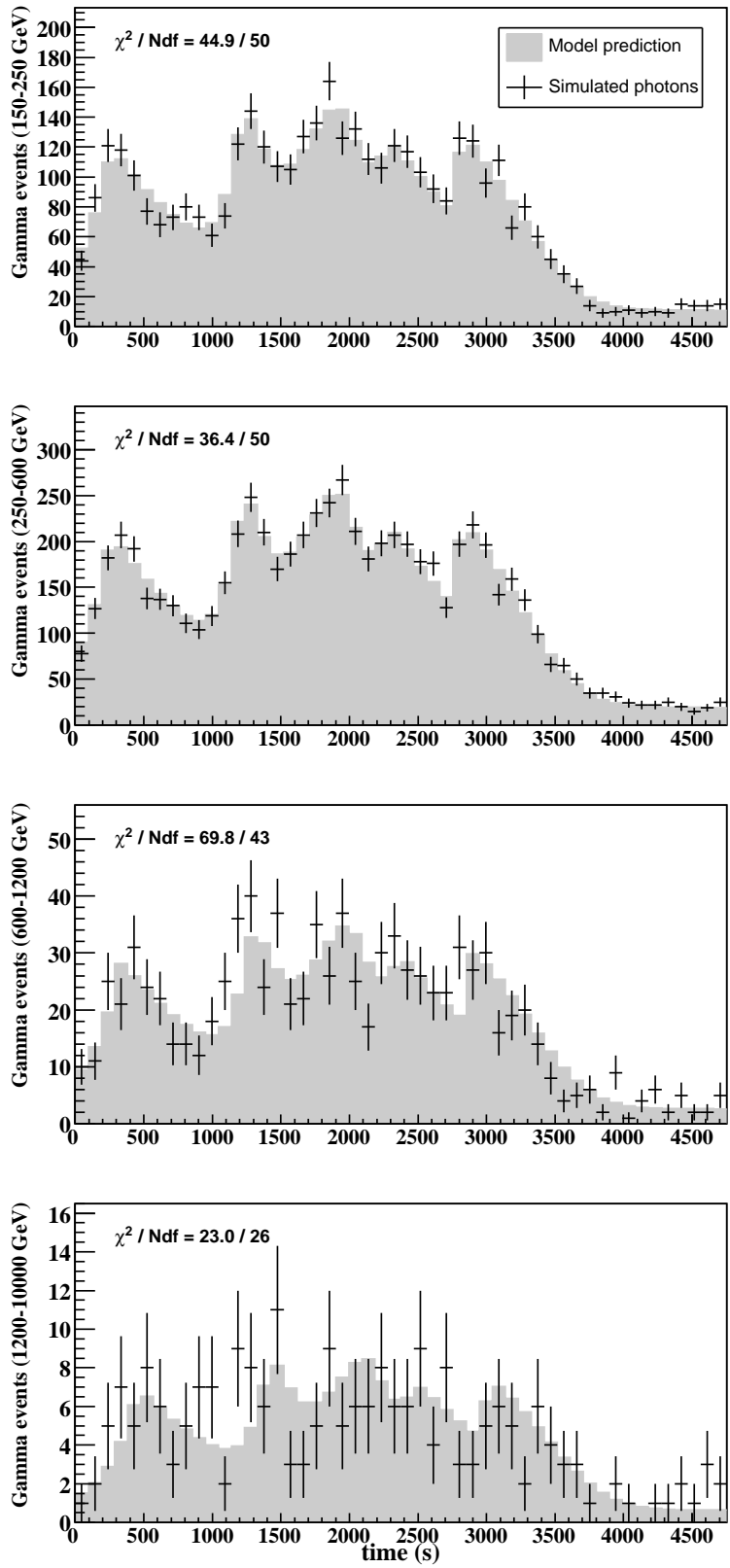}
\caption{Comparison of the binned HESS Monte Carlo data and
Likelihood function distributions to check the goodness-of-fit.
The energy intervals chosen are the same ones used in fig.
\ref{fig:MAGIC-binned} whereas the time bins here are of 95
seconds duration. The overall
$\chi^2$/Ndf is 174.1/168.} \label{fig:HESS-binned}
\end{figure}

Fig.~\ref{fig:HESS-binned} shows the comparison between the
binned Monte Carlo dataset and the Likelihood function for the
case in which a QG scale equal to the one observed in the MAGIC
data is assumed. In all tested cases, the fits do recover within
one sigma the $M_{QGn}$ value used in the simulations, and this
result demonstrates the reliability of our approach to analyze
complex flare structures.

Moreover, our fits predict that the H.E.S.S. data could provide a
sensitivity in the determination of any measurement of $M_{QGn}$
in the linear and in the quadratic assumptions, over a factor 6
better than the one obtained with the Markarian 501 flare observed
by MAGIC. By playing with our fits, we've been able to trace back
the origin of this improvement to three factors:

\begin{itemize}
\item{} A factor about three comes from the fact that the redshift
is about three times larger (0.116 compared with 0.034).

\item{} Another factor about two comes from the fact that the
gamma statistics is almost ten times larger (about 12 thousand
gammas compared with 1.4 thousand gammas). That would produce a
factor three smaller statistical uncertainty but, on the other
hand, the observed MAGIC spectrum is harder (more photons at
higher energies) than the H.E.S.S. one. In fact, it has been
checked by playing with the spectral indices that the photons with
the highest energies (above say 1 TeV) do play a crucial role in
improving the sensitivity.

\item{} Finally an additional small gain comes from the rich flare
structure although in general, the sensitivity is dominated by the
fastest risetime or falltime in the whole lightcurve.
\end{itemize}

We expect the H.E.S.S. collaboration to release soon the results
of their analysis of their flare and confirm or correct our
expectations.

\section{Conclusions}
\label{conclusions}

In this work we've presented a new approach for the analysis of
correlations between the arrival time and energy of photons that
can be applied to any observed set of photons coming from
astrophysical sources to search for any kind of intrinsic or
propagation energy-dependent delay effects.

This approach uses directly the energy and arrival time information
of each recorded photon and therefore does not require binning.
Unlike other unbinned approaches, the one presented should be able to
study the energy versus time correlations regardless on the
complexity of the flare structure and even with quite low photons
statistics. The method proposed is based upon a Likelihood estimator and
therefore should be statistically optimal to find the best
estimator for that correlation.

This analysis technique was already applied to the Markarian 501
flare observed by the MAGIC telescope in July 2005
\cite{magic-flare,magic-QG} and some details about the actual
implementation in that case, as well as of several tests performed
to verify the reliability, robustness and stability of the results
have been given. For that case, the goodness-of-fit has also been
discussed.

Finally, the application to a more complex
practical case such as the huge flare of PKS 2155-304 observed in
July 2006 by the H.E.S.S. collaboration has been discussed. Monte
Carlo simulations show that the method could be reliably used to
analyze that data.

\section*{Acknowledgements}
\label{Acknowledgements}

We're indebted to Robert Wagner, Adrian Biland and Rudy Bock from
the MAGIC Collaboration, and John Ellis, Nikolaos Mavromatos,
Dimitri Nanopoulos, Alexandre Sakharov and Edward
Sarkisyan-Grinbaum for many discussions and exchange of ideas
during the development of this method and during its application
to the MAGIC data. We're specially grateful to Adrian Biland for
providing the Monte Carlo samples for some of the test of the
application of this method to the MAGIC data and also for some
clever suggestions.

We want to thank David Paneque for providing us with all the
details of his analysis of the Markarian 501 flare and for being
so obstinate in claiming that there was some funny effect there.

We also want to thank the MAGIC-IFAE group for many discussions in
the early stages of the development of this method and specially
Javier Rico for his critical analysis about the interpretation of
its results. Moreover, we thank Rolf B\"uhler for suggesting a
modification in equation\:1 to make it more general.

Finally we want to thank all our colleagues of the MAGIC
Collaboration for making possible to produce such high quality
data as the one used for the application of our method to the
MAGIC data.


\begin{thebibliography}{00}

\bibitem{cheng}Cheng, K.S., Ho, C. and Ruderman, M., ApJ 300 (1986) 500.
\bibitem{harding}Daugherty, J.K. and Harding, A.K., ApJ 458 (1996) 278.
\bibitem{ghisellini}Ghisellini, G. and Tavecchio, MNRAS 386 (2008) 28.
\bibitem{bednarek07}Bednarek, W. and Wagner, R.M., Astron. Astrophys. 486 (2008) 679.
\bibitem{rovelli}Rovelli, C. and Smolin, L., Nucl. Phys. B 331 (1990) 80.
\bibitem{ellis92}Ellis, J.R., Mavromatos, N.E. and Nanopoulos, D.V., Phys. Lett. B 293 (1992) 37.
\bibitem{kostelecky}Kostelecky, V.A. and Potting, R., Phys. Lett. B. 381 (1996) 89.
\bibitem{amelino97}Amelino-Camelia, G., Ellis, J.R., Mavromatos, N.E. and Nanopoulos, D.V., Int J. Mod. Phys. A 12 (1997) 607.
\bibitem{gambini}Gambini, R. and Pullin, J., Phys. Rev. D 59 (1999) 124021.
\bibitem{klinkhamer07} Klinkhamer F.R., AIP Conf. Proc. 977 (2008) 181.
\bibitem{rovelli98}Rovelli, C., Living Reviews in Relativity 1 (1998) 1.
\bibitem{Ellis:1999yd}Ellis, J.R., Mavromatos, N.E. and Nanopoulos, D.V., ``Probing models of quantum space-time foam,'' arXiv:gr-qc/9909085.
\bibitem{sarkar02}Sarkar, S., Mod. Phys. Lett. A 17 (2002) 1025.
\bibitem{piran05}Piran, T., Lect. Notes Phys. 669 (2005) 5.
\bibitem{amelino98}Amelino-Camelia, G., Ellis, J.R., Mavromatos, N.E., Nanopoulos, D.V. and Sarkar, S., Nature 393 (1998) 763.
\bibitem{biller99}Biller, S.D. et al., Phys. Rev. Lett. 89 (1999) 2108.
\bibitem{kaaret99}Kaaret, P., Astron. Astrophys. 345 (1999) L32.
\bibitem{ellis06}Ellis, J.R. et al., Astropart. Phys. 25 (2006) 402.
\bibitem{blanch03}Blanch, O., L\'{o}pez, J. and Mart\'{i}nez, M., Astropart. Phys. 19 (2003) 245.
\bibitem{magic-flare}Albert, J. et al., ApJ 669 (2007) 862.
\bibitem{aharonian07}Aharonian, F. et al., ApJ 664 (2007) L71.
\bibitem{li} Li, T.P. et al., J. Astron. Astrophy. 4 (2004) 583.
\bibitem{edelson} Edelson, R.A. and Krolik, J.H., ApJ 333 (1988) 646.
\bibitem{mallat} Mallat, S., "A Wavelet Tour of Signal Processing",
Academic Press, San Diego, 1999.
\bibitem{ellis03}Ellis, J.R. et al., Astron. Astrophys. 402 (2003) 409.
\bibitem{lamon}Lamon, R., Produit, N. and Steiner, F., General Relativity and Gravitation 40 (2008) 1731.
\bibitem{bolmont} Bolmont, J. et al., ApJ 676 (2008) 532.
\bibitem{HESS-QG} Aharonian, F. et al. Phys. Rev. Lett. 101 (2008) 170402.
\bibitem{magic-QG}Albert, J. et al., Phys. Lett. B 668 (2008) 253.
\bibitem{norris96} Norris, J.P. et al., ApJ 662 (1996) 393.

\end{thebibliography}
\end{document}